\documentclass[conference]{IEEEtran} 
\IEEEoverridecommandlockouts
\usepackage{cite}
\usepackage{amsmath,amssymb,amsfonts}
\usepackage{algorithmic}
\usepackage{graphicx}
\usepackage{textcomp}
\usepackage{xcolor}
\usepackage{subfig}
\usepackage{multirow}
\def\BibTeX{{\rm B\kern-.05em{\sc i\kern-.025em b}\kern-.08em
    T\kern-.1667em\lower.7ex\hbox{E}\kern-.125emX}}
\begin{document}

\title{Exploiting Multi-domain Visual Information \\ for Fake News Detection}
\author{Anonymous}



\author{Peng Qi$^{1,2}$, Juan Cao$^{1,2}$, Tianyun Yang$^{1,2}$, Junbo Guo$^{1}$ and Jintao Li$^{1}$\\
\IEEEauthorblockA{
\textit{$^{1}$Institute of Computing Technology, Chinese Academy of Sciences, Beijing, China} \\
\textit{$^{2}$University of Chinese Academy of Sciences, Beijing, China}\\
\{qipeng,caojuan,yangtianyun,guojunbo,jtli\}@ict.ac.cn}
}

\maketitle

\begin{abstract}
The increasing popularity of social media promotes the proliferation of fake news.
With the development of multimedia technology, fake news attempts to utilize multimedia contents with images or videos to attract and mislead readers for rapid dissemination, which makes visual contents an important part of fake news.
Fake-news images, images attached in fake news posts, include not only fake images which are maliciously tampered but also real images which are wrongly used to represent irrelevant events.
Hence, how to fully exploit the inherent characteristics of fake-news images is an important but challenging problem for fake news detection.
In the real world, fake-news images may have significantly different characteristics from real-news images at both physical and semantic levels, which can be clearly reflected in the frequency and pixel domain, respectively.
Therefore, we propose a novel framework \underline{M}ulti-domain \underline{V}isual \underline{N}eural \underline{N}etwork (MVNN) to fuse the visual information of frequency and pixel domains for detecting fake news.
Specifically, we design a CNN-based network to automatically capture the complex patterns of fake-news images in the frequency domain; and utilize a multi-branch CNN-RNN model to extract visual features from different semantic levels in the pixel domain.
An attention mechanism is utilized to fuse the feature representations of frequency and pixel domains dynamically.
Extensive experiments conducted on a real-world dataset demonstrate that MVNN outperforms existing methods with at least 9.2\% in accuracy, and can help improve the performance of multimodal fake news detection by over 5.2\%. 
\end{abstract}


\section{Introduction}
Social media has become a major information platform, where people acquire the latest news and express their opinions freely. 
However, the convenience and openness of social media have also promoted the proliferation of \textit{fake news}, which caused many detrimental societal effects.
For instance,
during the month before the 2016 U.S. presidential election campaign, the American encountered between one and three fake stories in average from known publishers \cite{allcott2017social}, which inevitably misled the voters and influenced the election results. 
Hence, the automatic detection of fake news has become an urgent problem of great concern in recent years \cite{surveyKai2017Fake,surveykumar2018false,surveyzubiaga2018,surveywu2016mining}.

The development of multimedia technology promotes the evolution of self-media news from text-based posts to multimedia posts with images or videos, which provide better storytelling and attract more attention from readers. 
It is estimated that the average number of reposts for posts with images is 11 times larger than those without images\cite{jin2015TMM}.
Unfortunately, this advantage is also taken by fake news which usually contains misrepresented or even tampered images to attract and mislead readers for rapid dissemination. 
As a result, visual content has become an important part of fake news that cannot be neglected.

\begin{figure}
	\centering
	\subfloat[Tampered Images]{
		\label{Fig:manipulated}
		\includegraphics[width=0.35\textwidth,height=0.88in]{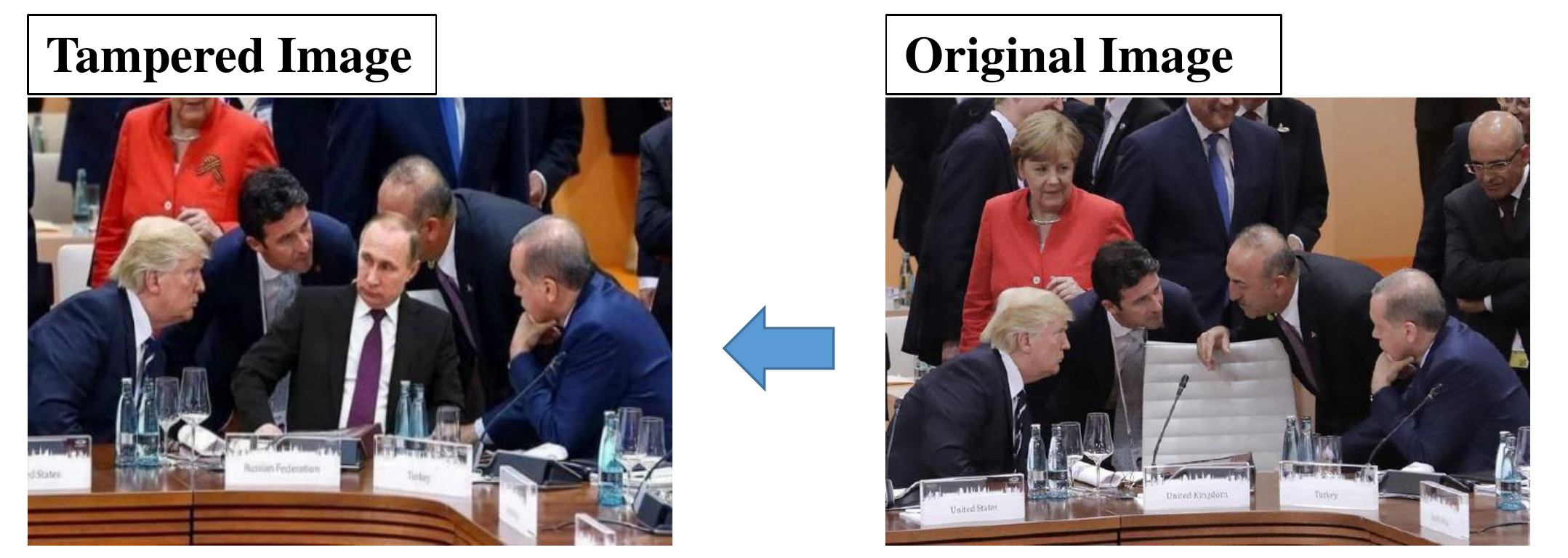}
	}\\
	\subfloat[Misleading Images]{
		\label{Fig:misleading}
		\includegraphics[width=0.45\textwidth,height=0.7in]{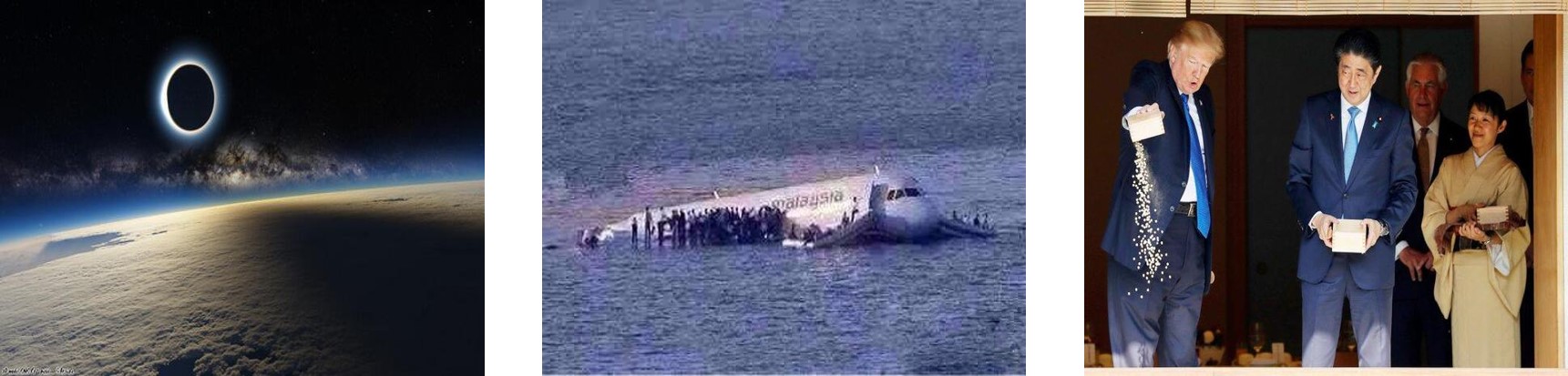}
	}\\
	\caption{Examples of fake-news images: 
		(a) A tampered image where Putin is spliced on the middle seat at G-20 to show that he is in the middle of an intense discussion between other world leaders. 
		(b) Three misleading images: the left one is an artwork which is reposted as a photo of solar eclipse (March 2015); the middle one is a real image captured in 2009 New York plane crash, but it is claimed to be the wrecked Malaysia Airlines MH370 in 2014; the right one is a real image used to misinterpret Trump's behavior while he is just imitating Abe.}
	\label{Fig:fakenews}
\end{figure}

\begin{figure*}
	[htbp]
	\centering
	\includegraphics[width=\textwidth]{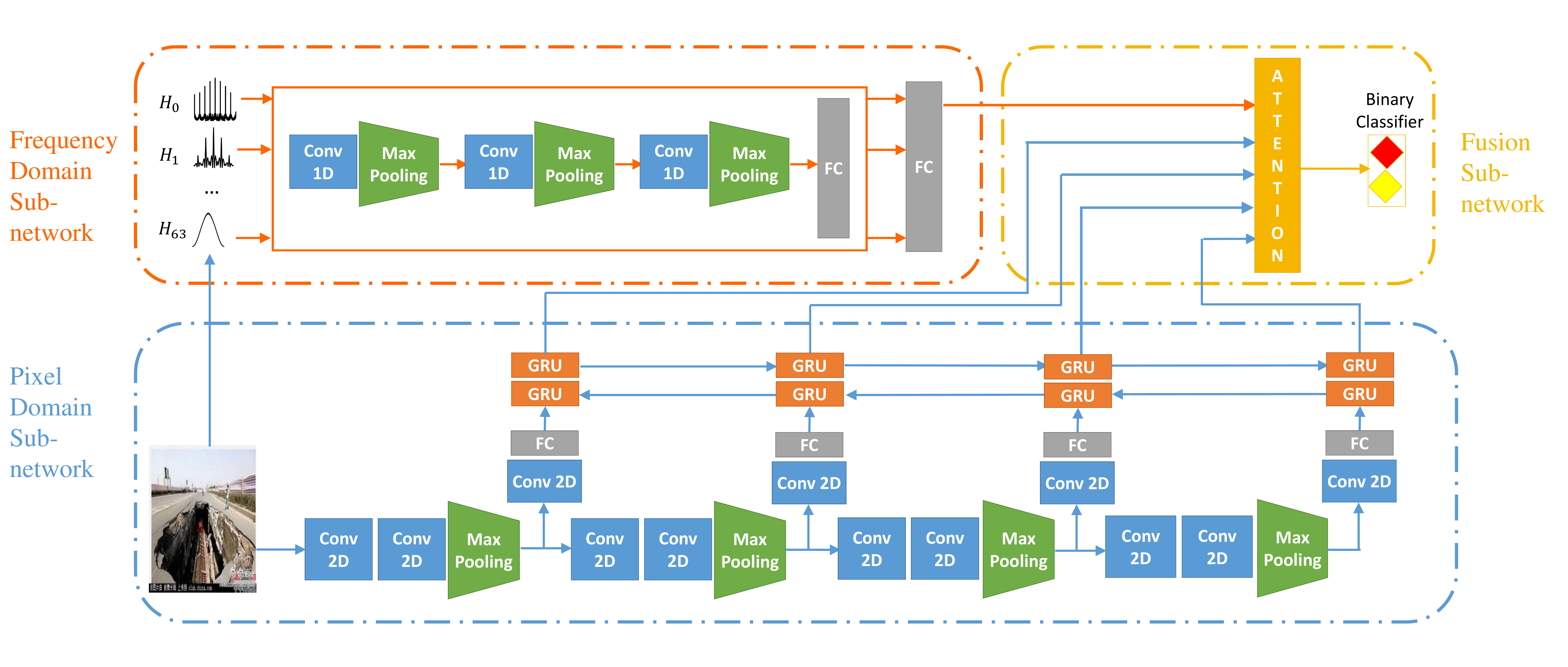}
	\caption{An illustration of the proposed Multi-domain Visual Neural Network (MVNN). It mainly consists of three components: a frequency domain sub-network, a pixel domain sub-network, and a fusion sub-network. The frequency domain sub-network first transforms the input image from pixel domain to frequency domain, and utilizes a CNN-based model to capture the physical characteristics of this image. The pixel domain sub-network employs a multi-branch CNN-RNN network to extract the features of different semantic levels of the input image. The fusion sub-network dynamically fuses feature vectors obtained from the frequency and pixel domain sub-network through an attention mechanism to classify the input image as a fake-news or real-news image.}
	\label{Fig:model}
\end{figure*}

According to existing research\cite{mediaeval16}, \textbf{\textit{fake-news images}}, images attached in fake news posts, include not only fake images which are maliciously tampered but also real images which are wrongly used to represent irrelevant events.
In detail, we broadly classify fake-news images into two categories: tampered images and misleading images.
\textbf{\textit{Tampered images}} mean fake-news images which have been digitally modified as Fig.~\ref{Fig:manipulated} shows, equal to \textit{fake images} in our common sense.
\textbf{\textit{Misleading images}} refer to fake-news images that haven't experienced any manipulation, but the content is misleading as described in Fig.~\ref{Fig:misleading}.
These misleading images usually derive from artworks or outdated images which are published on an early event \cite{outdatedimagessun2013apweb}.
Existing researches on fake news detection mainly focus on text content \cite{textcastillo2011information, statiswu2015whetherimageornot, ma2016ijcai,ma2019wwwgan} and social context \cite{socialwu2015false,socialshu2017exploiting}, while a few works utilize visual information to detect fake news.
Some works \cite{forensicsboididou2015certh, mediaeval16} extract forensics features to evaluate the authority of attached images, but these forensics features are mostly crafted manually for detecting specific manipulated traces, which are not applicable for misleading images.
Other works \cite{jin2017multimodal, eannwang2018, mvae} utilize pre-trained convolutional neural network like VGG19\cite{vgg19} to obtain general visual representations, which can hardly capture the semantic commonalities of fake-news images due to the lack of task-relevant information.
Therefore, how to fully exploit the inherent characteristics of fake-news images is an important but challenging problem for distinguishing fake news from real news by utilizing visual contents.

\begin{figure}
	\centering
	\subfloat[A fake-news image]{
		\label{frequencycase1}
		\includegraphics[height=0.9in]{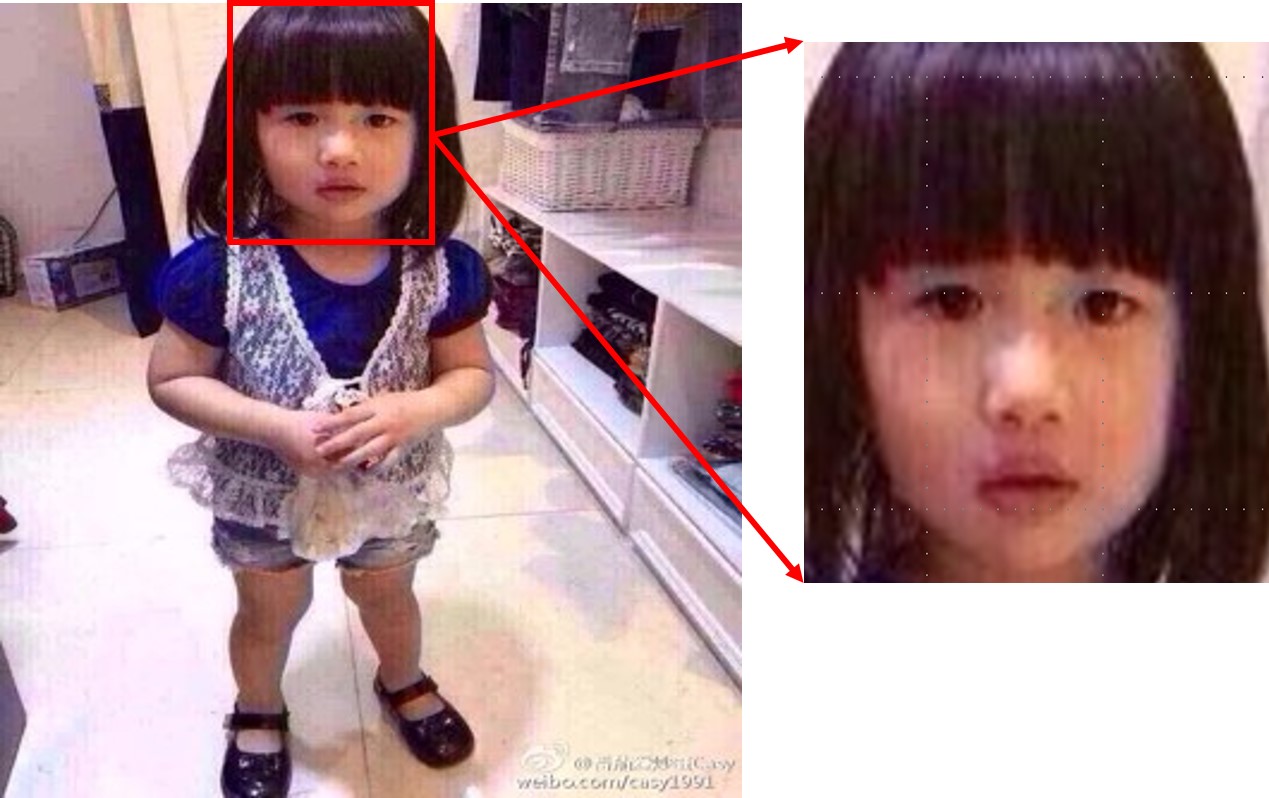}
	}
	\subfloat[A real-news image]{
		\label{frequencycase2}
		\includegraphics[height=0.9in]{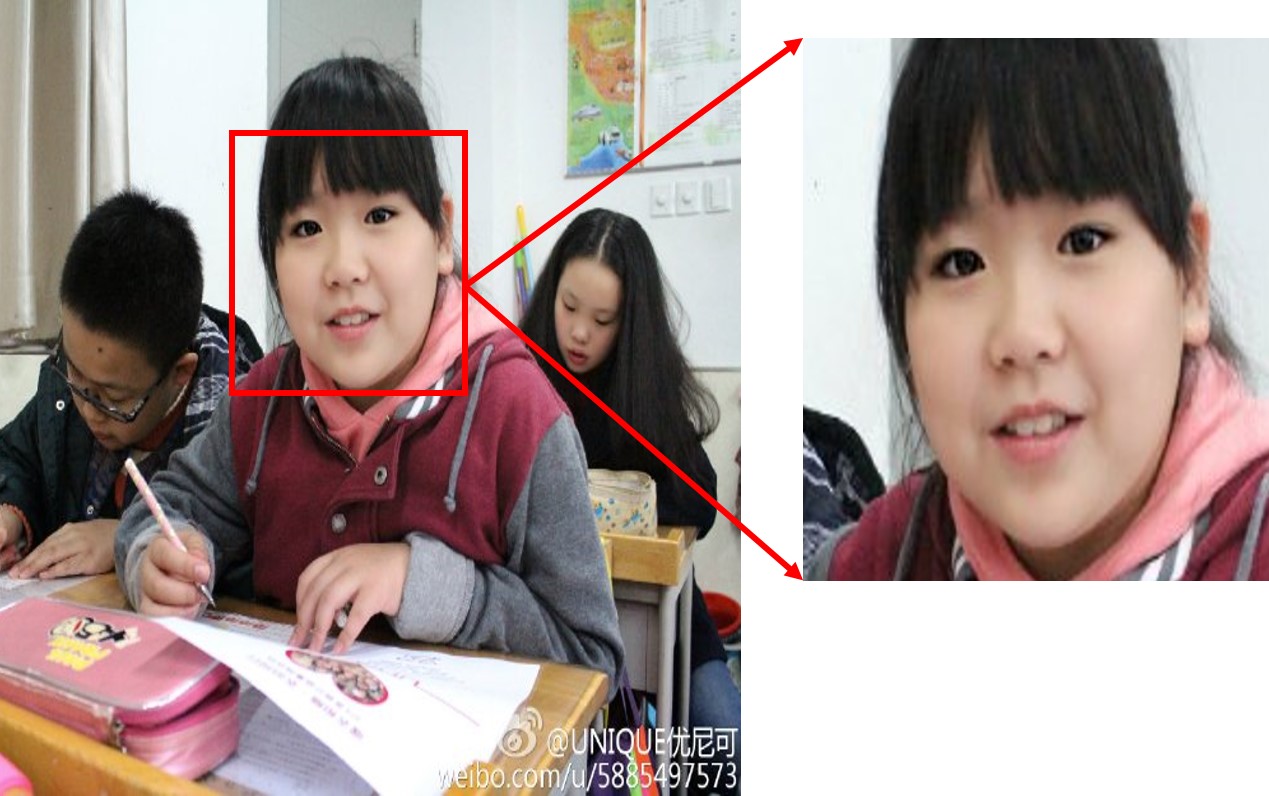}
	}\\
	\caption{Comparison of fake-news images and real-news images at the physical level. We observe that the fake-news image (a) has obvious block effect while the real-news image (b) is more clear. We enlarge the faces of the two images for better comparison.}
	\label{Fig:frequencycase}
\end{figure}

\begin{figure}
	\centering
	\subfloat[]{
		\label{pixelcase1}
		\includegraphics[height=1.5in]{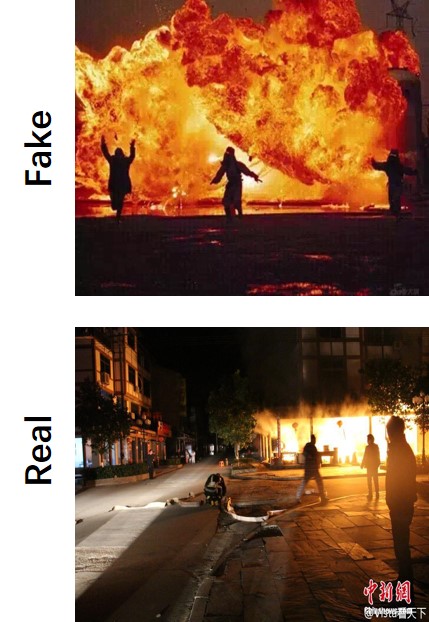}
	}
	\subfloat[]{
		\label{pixelcase1}
		\includegraphics[height=1.5in]{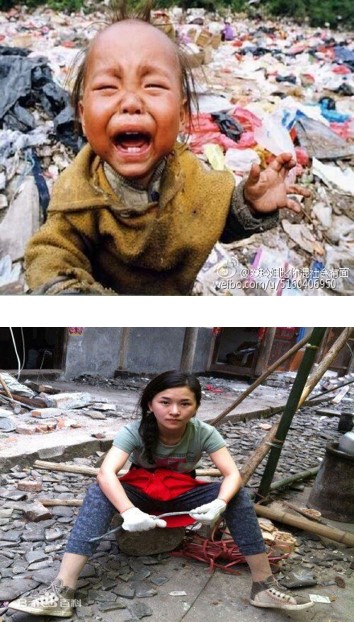}
	}
	\subfloat[]{
		\label{pixelcase3}
		\includegraphics[height=1.5in]{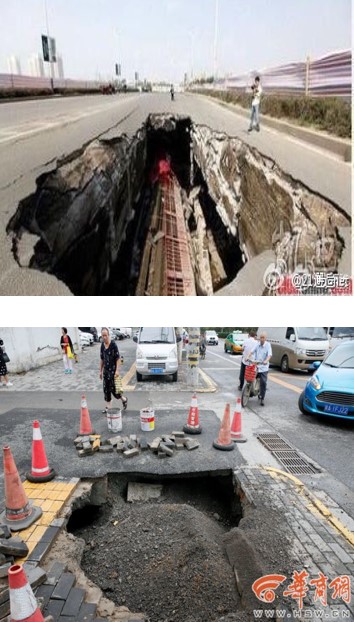}
	}\\
	\caption{Comparison of fake-news images and real-news images at the semantic level. We can find that fake-news images are more visually striking and emotional provocative than real-news images, even though they describe the same type of event such as fire (a), earthquake (b) and road collapse (c).}
	\label{Fig:pixelcase}
\end{figure}

In the real world, fake-news images may have significantly different characteristics from real-news images at both physical and semantic levels:

\textit{\textbf{ At the physical level}}, 
fake-news images might be of low quality, which can be clearly reflected in the \textbf{frequency domain}.
For example, 
after being uploaded to and downloaded from social platforms multiple times, 
misleading images usually have heavier re-compression artifacts than real-news images such as block effect as shown in Fig.~\ref{Fig:frequencycase}.
Besides, there are manipulation traces in tampered images inevitably.
Considering that re-compressed and tampered images often present periodicity in the frequency domain, which can be easily characterized by CNN which has the ability to capture spatial structure features, we design a CNN-based network to automatically capture the characteristics of fake-news images in the frequency domain (see the top part of Fig.~\ref{Fig:model}).

\textit{\textbf{At the semantic level},}
fake-news images also exhibit some distinct characteristics in the \textbf{pixel domain} (also known as spatial domain). 
Fake news publishers tend to utilize images to attract and mislead readers for rapid dissemination; thus fake-news images often show visual impacts\cite{jinarxiv2016image} and emotional provocations\cite{bookonrumours, surveyKai2017Fake} just as Fig.~\ref{Fig:pixelcase} shows. 
These characteristics have been proven related to many visual factors from low-level to high-level \cite{emotionstimuli}; thus we build a multi-branch CNN-RNN network to extract features of different semantic levels (see the bottom part of Fig.~\ref{Fig:model}), for fully capturing the characteristics of fake-news images in the pixel domain.

We have shown that the visual features at both physical and semantic levels are important for detecting fake-news images; thus fusing the visual information of the frequency and pixel domains has the potential to improve the performance of fake news detection.
Intuitively, 
not all features contribute equally to the task of fake news detection, which means some visual features play more important roles than others in evaluating whether a given image is fake-news or real-news image.
Therefore, we employ an attention mechanism \cite{attention} to fuse these visual features from different domains dynamically.

To sum up, 
we propose a \underline{M}ulti-domain \underline{V}isual \underline{N}eural \underline{N}etwork (MVNN) framework (see Fig.~\ref{Fig:model}), which 
can learn effective visual representations by combining the information of frequency and pixel domains for fake news detection. 
The proposed model consists of three main components: 
a frequency domain sub-network and a pixel domain sub-network used to capture the characteristics of fake-news images at physical and semantic levels respectively, and a fusion sub-network to fuse these features dynamically.
In summary, the contributions of this paper are three folds:
\begin{enumerate}
	
	\item 
	To the best of our knowledge, we are the first to utilize multi-domain visual information for fake news detection, which captures the inherent characteristics of fake-news images at both physical and semantic levels. 
	
	\item  We propose a novel framework MVNN, which exploits an end-to-end neural network to learn representations of frequency and pixel domains simultaneously and effectively fuse them. 
	
	\item We conduct extensive experiments on a real-world dataset to validate the effectiveness of the proposed model.
	The results demonstrate that our model is much better than existing methods, and 
	the visual representations learned by our model can help improve the performance of multimodal fake news detection by a large margin. 
	Besides, we show that the information of frequency and pixel domains are complementary for detecting fake news.     
\end{enumerate}

\section{Related Work} 
In the task of fake news detection, the main challenge is how to utilize information from different modalities to distinguish fake news posts and real news posts.
Most of existing approaches focus on text content \cite{textcastillo2011information, statiswu2015whetherimageornot, ma2016ijcai,ma2019wwwgan} and social context \cite{socialwu2015false,socialshu2017exploiting} which refers to information generated during the news propagation process on social networks.
%
%
Recently, visual information has been shown to be an important indicator for fake news detection\cite{jin2015TMM, surveyKai2017Fake}.
With the popularity of multimedia content, researchers begin to incorporate visual information to detect fake news.

Some early works use basic statistical features about attached images to help classify fake news posts, such as the number of attached images\cite{statisyang2012automatic, statiswu2015whetherimageornot}, image popularity and image type\cite{jin2015TMM}.
However, these statistical features are so basic that they can hardly represent complex distributions of visual contents in fake news.

Visual forensics features are generally used for image manipulation detection.
To evaluate the authority of attached images, 
some works extract visual forensics features, such as block artifact grids (BAG), to assist fake news detection.
For example, the Verifying Multimedia Use task on 2015 \cite{mediaeval15} and 2016 MediaEval benchmark\cite{mediaeval16} provide seven types of image forensics features to help detect manipulated and misleading use of web multimedia content.
Based on these provided forensics features, \cite{forensicsboididou2015certh} extract advanced forensics features and combine them with post-based and user-based features in a semi-supervised learning scheme to tackle the news verification problem.
However, most of these forensics features are crafted manually for detecting specific manipulated traces, which are not applicable for the real images attached in fake news.
Besides, these hand-crafted features are labor-expensive and limited to learn complicated patterns, causing the poor generalization performance on the task of fake news detection.

Inspired by the power of CNN, most existing works based on multimedia contents use a pre-trained deep CNN like VGG19\cite{vgg19} to obtain general visual representations and fuse them with text information.
Specifically, 
\cite{jin2017multimodal} first incorporates multimodal contents on social networks via deep neural networks to solve fake news detection problem;
\cite{eannwang2018} proposes an end-to-end event adversarial neural network to detect newly-emerged fake news events based on multi-modal features;
\cite{mvae} presents a novel approach to learn a shared representation of multimodal information for fake news detection.
However, these works focus on how to fuse the information of different modalities, ignoring effectively modeling visual contents.
These visual features they adopted are too general to reflect the intrinsic characteristics of fake-news images due to the lack of task-relevant information, which degrades the performance of visual contents in fake news detection.

To overcome these limitations of existing works, we propose a novel deep learning network to model visual contents at both physical and semantic levels for the task of fake news detection. 


\section{Problem Formulation}
Fake news is defined as news articles that are intentionally and verifiably false and could mislead readers\cite{fakenewsdefinition}, which has been widely adopted in recent studies\cite{surveyKai2017Fake}.
Different from the traditional definition, 
in the context of microblog, fake news aims at news posts that are published on social media by users which are usually less than 140 characters instead of news articles.
Formally, we state this definition as follows,

\textit{Definition 1: \textbf{Fake news:} In the context of microblog, a piece of fake news is a news post that is intentionally and verifiably false. }

\textit{Definition 2: \textbf{Fake-news images:} A fake-news image is an image attached in fake news.}

The problem addressed in this paper is how to utilize the visual content to identify a news post as real or fake, which equals to classify a given image as fake-news image or real-news image. 
Consequently, we formally definite the studied problem as follows:

\textit{\textbf{Problem 1}: Given a set of news posts $\mathcal{X}=\left\{x_{1}, x_{2}, \dots, x_{m}\right\}$, corresponding images $\mathcal{I}=\left\{i_{1}, i_{2}, \dots, i_{m}\right\}$, and labels $\mathcal{Y}=\left\{y_{1}, y_{2}, \dots, y_{m}\right\}$, learn a classifier $f$ that can utilize the corresponding image to classify whether a given post is fake news ($y_{t}=1$) or real news ($y_{t}=0$) , i.e., $\hat{y_t} = f(i_t)$. }

\section{Methodology} 
In this section, we first make an overview of the proposed Multi-domain Visual Neural Network (MVNN), and then introduce the three major components in detail.

\subsection{Model Overview}
The goal of the proposed MVNN model is utilizing the visual information in frequency and pixel domains to evaluate whether the given image is a fake-news or real-news image. 
As shown in Fig.~\ref{Fig:model}, MVNN includes three major components: a frequency domain sub-network, a pixel domain sub-network and a fusion sub-network.
For an input image, we feed it into the frequency and pixel domain sub-network to obtain features at physical and semantic levels respectively.
The fusion sub-network takes these learned features as input and learns a final visual representation of this image to predict whether this image is a fake-news or real-news image. 

\subsection{Frequency Domain Sub-network}
\begin{figure}
	\centering
	\includegraphics[width=0.5\textwidth]{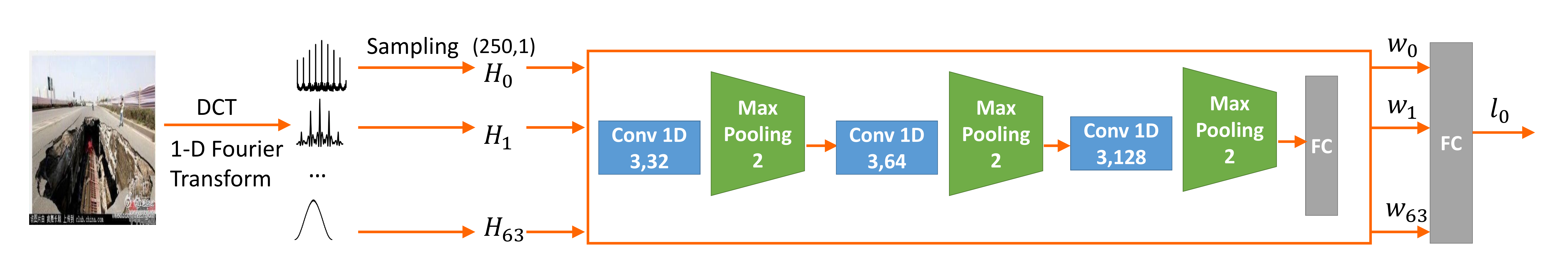}
	\caption{
	The detailed architecture of the frequency domain sub-network.
	For an input image, we first transform it from pixel domain to frequency domain, and design a CNN-based network to obtain its feature representation in the frequency domain.}
	\label{Fig:frequencymodel}
\end{figure}

The physical features of images can reflect their originality to a certain extent which is helpful for detecting fake news; thus we design the frequency domain sub-network to extract features of the input image at the physical level. 
According to existing studies about image forensics \cite{dctpr09,dcttifs11,dcttifs12,dcteurasip16}, discrete cosine transform (DCT) has been widely utilized for capturing the tampered and re-compressed architects, so we employ DCT on the input image to transform it from pixel domain to frequency domain. 
Considering that tampered images and re-compressed images often present periodicity in the frequency domain, which can be easily characterized by CNN with the ability to capture spatial structure characteristics; thus we design a CNN-based network to capture the characteristics of fake-news images in the frequency domain. 
The detailed architecture of this part is shown in Fig.~\ref{Fig:frequencymodel}. 

For an input image, we first employ block DCT on it to obtain 64 histograms of DCT coefficients corresponding to 64 frequencies.
Following the process of \cite{dcttifs11}, we then carry 1-D Fourier transform on these DCT coefficient histograms to enhance the effect of CNN.
Considering that CNN needs an input of a fixed size, we sample these histograms and obtain 64 250-dimensional vectors, which can be represented as $ \{H_0, H_1,...H_{63} \}$.
After being pre-processed, each input vector is fed to the shared CNN network to obtain corresponding feature representation $ \{w_0, w_1,...w_{63} \}$.
The CNN network consists of three convolutional blocks and a fully connected layer, and each convolutional block is composed of a one-dimensional convolutional layer followed by a max-pooling layer.   
To accelerate the convergence of the model, we set the number of filters in the convolution layers to be incremental.
Existing works about image forensics usually only consider coefficients of a part of frequencies.
However, we find that all frequencies contribute to the task of fake news detection,
hence we fuse the feature vectors of all frequencies by concatenating and feed the final feature representation $l_0$ to the fusion sub-network.
Specifically, we experiment with different fusing methods, and results show that concatenating performs best in this task.

\subsection{Pixel Domain Sub-network}
\begin{figure}[b]
	\centering
	\includegraphics[width=0.5\textwidth]{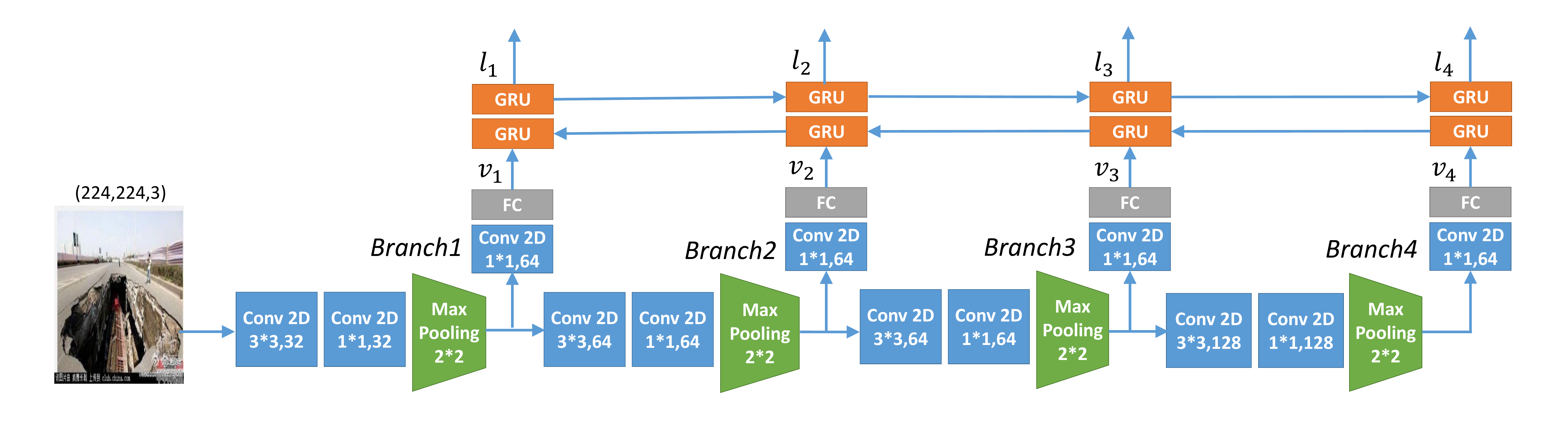}
	\caption{
	The detailed architecture of the pixel domain sub-network.
	For an input image, we utilize a multi-branch CNN-RNN network to extract its features of different semantic levels in the pixel domain.}
	\label{Fig:pixelmodel}
\end{figure}

We have shown that fake-news images have different characteristics from real-news images at the semantic level, indicating that the semantic features are important in detecting fake news.
Therefore, we design the pixel domain sub-network for extracting visual features of the input image at the semantic level, of which the detailed architecture is shown in Fig.~\ref{Fig:pixelmodel}. 

CNN learns high-level semantic representations through layer-by-layer abstraction from local to global view, of which the earlier layers prefer low-level features such as color, line and shape, while the later layers tend to focus on the objects.
In the process of abstraction, low-level features will inevitably suffer some losses, which further suggests that the bottom and intermediate layer in CNN can provide complementary information for the top layer.
Many works have proven integrating features from different layers can help achieve better performance than only using high-level features for some tasks such as salient object detection and image emotion classification \cite{aestheticsyang2018understanding, emotioncnnrnn,amulet}.
We have shown that fake-news images usually show some visual impacts and emotion provocation, which have been proven related to many visual factors from low-level to high-level\cite{emotionstimuli}.
Therefore, to fully capture the semantic characteristics of fake-news images, we build a multi-branch CNN network to extract multiple levels of features and utilize a bidirectional GRU (Bi-GRU) network to model the sequential dependencies between these features.

As Fig.~\ref{Fig:pixelmodel} described, the CNN network mainly consists of four blocks composed of a $3\times3$ and $1 \times 1$ convolution layer and a max-pooling layer. 
The input image is fed to the CNN, and features extracted from four branches will pass through a $1 \times 1$ convolutional layer and a fully connected layer,
obtaining the corresponding feature vectors ${v_t}, t\in[1,4]$ .
These features represent different parts of images, such as line, color, texture, and object, which characterize different levels of features from local to global view. 
Motivated by Inception module used in GoogLeNet\cite{googLeNet}, we use $1 \times 1$ convolution layer to reduce the dimension and increase the model's representative power because it increases the non-linear activation and promotes the information fusion of different channels.

Intuitively, there are strong dependencies between different levels of features. For example, middle-level features such as textures, consist of low-level features such as lines, and meanwhile compose high-level features such as objects. Therefore, we utilize GRU to model these dependencies between low-level and high-level features.
Specifically, we model these features from different levels as a sequence $ V = \{v_t\}, t\in[1,4] $, where $ v_t $ represents the visual features extracted from t-th branch in the CNN network described in Fig.~\ref{Fig:pixelmodel}.
Then the entire pipeline in GRU at time step t can be presented as follows:
\begin{equation}
r_{t}=\sigma(W_r [v_t, h_{t-1}] + b_{r})
\end{equation}
\begin{equation}z_{t}=\sigma(W_z [v_t, h_{t-1}] + b_{z})\end{equation}
\begin{equation}\tilde{h}_{t}=\tanh (W_{\tilde{h}} [v_t, r_{t} \odot h_{t-1}] + b_{\tilde{h}})\end{equation}
\begin{equation}h_{t}=\left(1-z_{t}\right) \odot h_{t-1}+ z_{t} \odot \tilde{h}_{t}\end{equation}
where $r_{t}, z_{t}, \tilde{h}_{t}, h_{t}$ are the reset gate, update gate, hidden candidate, and hidden state respectively. $ W$ are the weight matrices and $b$ are the bias terms. Besides, $\sigma$ stands for the sigmoid function and $\odot$ represents the element-wise multiplication.

Considering that the dependencies among different levels of features can be estimated from both local to global view and global to local view; we utilize the bidirectional GRU to model the relations from two different views. 
The Bi-GRU contains the forward GRU $ \overrightarrow{f} $ which reads from $ v_1 $ to $ v_4 $, and backward GRU $ \overleftarrow{f} $ which reads from $ v_4 $ to $ v_1 $:
\begin{equation} \overrightarrow{l_t} = \overrightarrow{GRU} (v_t), t \in [1,4]\end{equation}
\begin{equation}\overleftarrow{l_t} = \overleftarrow{GRU} (v_t), t \in [1,4]\end{equation}
For each timestep $t$, we obtain the hidden representations by concatenating the forward hidden state $\overrightarrow{l_t}$ and the backward hidden state $\overleftarrow{l_t}$, i.e., $ l_t = [\overrightarrow{l_t}, \overleftarrow{l_t}]$, which compose the final semantic feature representation $ L = \{l_t\}, t\in[1,4] $.

\subsection{Fusion Sub-network}
We assume that the physical and semantic features of images are complementary in detecting fake news; thus we propose the fusion sub-network to fuse these features. 
Specifically, we utilize the output vector of the frequency domain sub-network $l_0$ and pixel domain sub-network $\{l_1, l_2, l_3, l_4\}$ to make predictions. 
$\{l_1, l_2, l_3, l_4\}$ represent visual features from different semantic levels while $l_0$ represents visual feature in the physical level.
Intuitively, 
not all features contribute equally to the task of fake news detection, which means some visual features play a more important role than others in evaluating whether a given image is a fake-news or real-news image.
For instance, for some tampered images which presents obvious tampering traces, the physical features perform better than semantic features; 
for some misleading images which have not experienced severe re-compressions, the semantic features are more effective.
Therefore, we highlight those valuable features via attention mechanism, and the enhanced image representation is computed as follows:
\begin{equation}\mathcal{F}(l_i)=v^{T} \tanh (W_f l_i+b_f), i\in[0,4] \end{equation}
\begin{equation}\alpha_i=\frac{\exp \left(\mathcal{F}\left(l_i\right)\right)}{\sum_{i} \exp \left(\mathcal{F}\left(l_i\right)\right)}\end{equation}
\begin{equation}u=\sum_{i} \alpha_i l_i\end{equation}
where $W_f$ denotes the weight matrix and $b_f$ is the bias term, $v^{T}$ denotes a transposed weight vector and $\mathcal{F}$ is the score function which measures the significance of each feature vector. Afterward, we obtain the normalized weight of i-th feature vector $\alpha_i$ via a softmax function and compute the advanced representation of the input image as a weighted sum of different feature vectors. The vector $v$ is randomly initialized and jointly learned
during the training process.

Till now, we have obtained a high-level representation $u$ of the input image, which models the characteristics of this image at both physical and semantic levels. 
We use a fully connected layer with softmax activation to project this vector into the target space of two classes: fake-news image and real-news image, and gain the probability distributions:
\begin{equation}p=\operatorname{softmax} \left(W_{c} u + b_{c}\right)\end{equation}
In the proposed model MVNN, we define the loss function as cross-entropy error between the predicted probability distribution and the ground truth:
\begin{equation}L=-\sum\left[y \log p +\left(1-y\right) \log \left(1-p\right)\right]\end{equation}
where $y$ is the ground truth with 1 representing fake-news image and 0 representing real-news image and $p$ denotes the predicted probability of being fake-news image.

\section{Experiments} 

In this section, we conduct experiments on a real-world dataset to evaluate the effectiveness of the proposed model MVNN. Specifically, we aim to answer the following evaluation questions:
\begin{itemize}
	\item \textbf{EQ1}: Is MVNN able to improve the performance of fake news detection based on visual modality?
	\item \textbf{EQ2}: How effective are different domains and other network components: attention, Bi-GRU and branches in the pixel domain sub-network, in improving the performance of MVNN?
	\item \textbf{EQ3}: Can MVNN help improve the performance of multimodal fake news detection? 
\end{itemize}
We first introduce the dataset we used, and some representative methods which models the visual contents for fake news detection as baselines; next we describe the implementation details of the proposed MVNN model.
And then we compare MVNN with those baselines to answer \textbf{EQ1}, and we investigate the ablation study to answer \textbf{EQ2}, which includes quantitative and qualitative analyses.
To answer \textbf{EQ3}, we further compare MVNN with baselines on multimodal fake news detection.
Last, we study some typical cases to vividly illustrate the importance of multiple domains for fake news detection.

\subsection{Dataset}

Considering that fake news detection based on multimedia content is a pretty new task, there are a few standard multimedia fake news datasets available. 
The two most widely used datasets are Twitter dataset presented in the MediaEval Verifying Multimedia Use benchmark\cite{mediaeval16}, and Weibo dataset built in \cite{jin2017multimodal}.
However, 
there are a lot of duplicate images in the Twitter dataset, causing the amount of distinctive images less than 500, which makes Twitter dataset too small to support the training of the proposed model.  
Therefore, in this paper, we conduct our experiments on Weibo dataset alone to evaluate the effectiveness of the proposed model.

In Weibo dataset, 
the fake news are crawled from May, 2012 to January, 2016 and verified by the official rumor debunking system of Weibo\footnote{https://service.account.weibo.com/}, which actually serves as an reputable source to collect fake news posts in literature.
The real news are collected during the same period as the fake news from Weibo and are verified by Xinhua News Agency, an authoritative news agency in China. 
According to \cite{jin2017multimodal},
duplicated and very small images have been removed for ensuring the quality of entire dataset.
To ensure that each post corresponds to an image, text-only posts are removed and only one image is saved for posts with multiple illustrations.
In total, this dataset includes 4749 fake news posts and 4779 real news posts with corresponding images.
In our experiment, we first use K-means algorithm to cluster all news posts into 200 clusters, and split the whole dataset into training set, validation set and testing set to ensure that there is no event overlap between these sets, which further prevents the model from overfitting on event topics.
The training, validation and testing sets contain a number of posts approximately with a ratio of 7:1:2 as in \cite{eannwang2018}.

\subsection{Baselines}
\label{baselines}
To validate the effectiveness of MVNN, we choose several representative methods which are used to model the visual contents for fake news detection as baselines:
\begin{itemize}
	\item \textbf{Forensics Features (FF)+LR}: Reference \cite{forensicsboididou2015certh} proposes new image forensics features and evaluate the effectiveness of these features in detecting fake news. We adopt logistic regression (LR) algorithm to take use of these features to make classifications.
	\item \textbf{Pre-trained VGG}: Pre-trained VGG are widely used as a feature extractor in existing studies about multimodal fake news detection \cite{jin2017multimodal, eannwang2018, mvae}. Specifically, these works use the output of the last layer of a 19-layer VGGNet pre-trained on ImageNet, of which the feature dimension is 4096. Similarly, we use LR to classify these features.
	\item \textbf{Fine-tuned VGG}: It has become a common practice that fine tuning a pre-trained model on a task-relevant dataset for achieving a promising performance on specific tasks. Hence, in addition to Pre-trained VGG, we also use Fine-tuned VGG as a compared experiment. 
	In particular, we use the training data to fine tune pre-trained VGG19, and use the same method as Pre-trained VGG to extract features and make classifications. 
	\item \textbf{ConvAE}: AutoEncoder (AE) is a kind of artificial neural network used to learn efficient data codings in an unsupervised manner. Convolutional AutoEncoder (ConvAE) \cite{ConvAE} extends the AE framework through utilizing convolutional layer to compose the encoder and decoder for obtaining better understanding of images than plain fully connected layers.
	Considering that only a few methods are used to model the visual contents for fake news detection, we use ConvAE as an additional comparison method to show the full advantage of the proposed model. 
	In order to avoid any bias, we design the encoder to be the same structure as the pixel domain sub-network without branches.
	After pre-training the ConvAE, we use the encoder to extract features and use LR to make classifications.
\end{itemize}

\subsection{Implementation Details}
In this section, we introduce the implementation details of the proposed model for reproducibility, including parameter setting and training strategy.  
In the frequency domain sub-network, the two fully connected layers contain 16 and 64 neurons respectively, and we add a dropout layer with a rate of 0.4 after the last fully connected layer to avoid overfitting.
In the pixel domain sub-network, the number of hidden units of the GRU and fully connected layer is 32 and 64, and each fully connected layer is followed by a dropout layer with a rate of 0.5. 
The number of branches is also a hyperparameter,
and we find that 4 performs best after a lot of experiments.
The frequency and pixel domain sub-networks are pre-trained and further fine tuned during the jointly training process.
When pre-training the pixel domain sub-network, we use data augmentation strategy to improve its generalization performance.   
In the whole network, we add a lot of batch normalization layers to accelerate its convergence, and we set the batch size to be 64.
The model is trained for 300 epochs with early stopping to prevent overfitting.
We use ReLU as the non-linear activation function, and use Adam algorithm to optimize the loss function.

\subsection{Performance Comparison}
In this subsection, we conduct some experiments to compare the performance of MVNN and existing baselines described in Section \ref{baselines} (\textbf{EQ1}).
We use the accuracy, precision, recall and F1 score of the fake-news image class as evaluation metrics. 

\begin{table}
	\caption{The performance comparison of fake news detection in single visual modality.}
	\begin{center}
		\begin{tabular}{lcccc}
			\hline
			\textbf{Method}&\textbf{Accuracy}&\textbf{Precision}&\textbf{Recall}&\textbf{F1}\\
			\hline
			FF+LR &  0.650&0.612&0.579&0.595\\
			Pre-trained VGG & 0.721&0.669&0.738&0.702	\\
			Fine-tuned VGG & 0.754&0.74&0.689&0.714\\
			ConvAE & 0.734&0.685&0.744&0.713\\
			\textbf{MVNN} & \textbf{0.846}&\textbf{0.809}&\textbf{0.857}&\textbf{0.832}\\
			\hline
		\end{tabular}
		\label{tab:visual}
	\end{center}
\end{table}

From Tab.~\ref{tab:visual}, we can see that:
1) MVNN is much better than other baselines, which validates that MVNN can effectively capture the intrinsic characteristics of fake-news images.
Specifically, MVNN achieves an accuracy of 84.6\%, outperforming existing approaches by 9.2\% at least.
2) Intuitively, Fine-tuned VGG performs better than Pre-trained VGG, indicating that the learned features are more relevant to the task of fake news detection after fine-tuning the model on the fake news dataset.
3) The performance of ConvAE is slightly better than Pre-trained VGG. This demonstrates that ConvAE has the ability of understanding universal semantics of images, which is similar to models pre-trained in a supervised manner.
4) The performance of FF+LR is the worst among these compared methods, because the information captured by these forensics features is very limited.

\subsection{Ablation Study}
In this subsection, we aim to evaluate the effectiveness of multiple domains and other network components in improving the performance of MVNN \textbf{(EQ2)} from the quantitative and qualitative perspectives.

\noindent
\textbf{Quantitative Analysis}

To intuitively illustrate the effectiveness of different domains and other network components, we design several internal models for comparison, which are simplified variations of MVNN by removing certain components:
\begin{itemize}
	\item \textbf{w/o frequency domain (fd):} The frequency domain sub-network is removed from MVNN while the attention mechanism is saved for fusing features of the pixel domain sub-network.
	\item \textbf{w/o pixel domain (pd):} We removed the pixel domain sub-network and associated attention mechanism. The remaining structure is the frequency domain sub-network and the binary classifier.
	\item \textbf{w/o attention:} MVNN without the attention mechanism. Features from the frequency and pixel domain sub-networks are concatenated to make classification.
	\item \textbf{w/o Bi-GRU:} MVNN without the Bi-GRU in the pixel domain sub-network. 
	\item \textbf{w/o branches:} In addition to remove the Bi-GRU, we also remove the branches in the pixel domain sub-network. The input image is fed into the pixel domain sub-network and obtains an output vector from the last block, which is further used to make classification.
\end{itemize}

\begin{table}[t]
	\caption{Ablation study of MVNN.}
	\begin{center}
		\begin{tabular}{lcccc}
			\hline
			\textbf{Method}&\textbf{Accuracy}&\textbf{Precision}&\textbf{Recall}&\textbf{F1}\\
			\hline
					\textbf{MVNN} & \textbf{0.846} & \textbf{0.809} & \textbf{0.857}& \textbf{0.832}\\
					w/o frequency domain & 0.794  & 0.792 & 0.728 & 0.758\\
					w/o pixel domain & 0.737 & 0.698 & 0.717 & 0.708\\
					
					w/o attention & 0.827 & 0.778 & 0.853 & 0.814 \\
					w/o Bi-GRU & 0.828 & 0.772 & 0.841 & 0.805\\
					w/o branches & 0.803 & 0.752 & 0.830 & 0.789 \\
			\hline
		\end{tabular}
		\label{tab:ablation}
	\end{center}
\end{table}


The results of the ablation study are reported in Tab.~\ref{tab:ablation}.
We have the following observations: 

\subsubsection{Multiple domains}  
The frequency and pixel domain both play important roles for fake news detection, especially the pixel domain. Specifically, the accuracy drops by 5.2\% and 10.9\% without the frequency and pixel domain sub-network correspondingly, which further demonstrates that physical and semantic visual information are both important for detecting fake news. 
Besides, the performance of MVNN without pixel domain is obviously worse than frequency domain. It illustrates that pixel domain plays a major role in fake news detection while frequency domain is auxiliary. 

\begin{figure*}[htbp]
	\centering
	\subfloat[]{
		\label{Fig:tsne-frequency}
		\begin{minipage}[t]{0.3\textwidth}
			\includegraphics[height=1.2in]{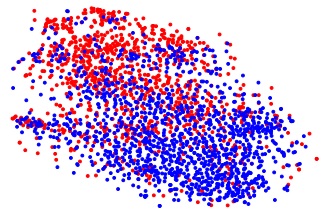}
		\end{minipage}
	}
	\subfloat[]{
		\label{Fig:tsne-pixel}
		\begin{minipage}[t]{0.3\textwidth}
			\includegraphics[height=1.2in]{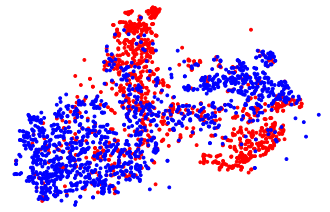}
		\end{minipage}
	}
	\subfloat[]{
		\label{Fig:tsne-fusion}
		\begin{minipage}[t]{0.3\textwidth}
			\includegraphics[height=1.2in]{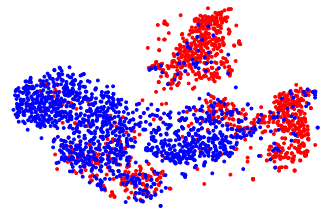}
		\end{minipage}
	}\\
	\caption{Visualization of learned latent visual feature representations on the testing data. 
	Dots in red and blue represent fake-news images and real-news images, respectively.
	(a) Frequency domain sub-network. (b) Pixel domain sub-network. (c) MVNN.}
	\label{Fig:tsne}
\end{figure*}

\subsubsection{Network Components} 
Other than multiple domains, other network components: attention, Bi-GRU and branches in pixel domain sub-network, 
are all important for achieving the best performance by MVNN. If we remove one or several of them, the performance would drop by a certain degree. 
Specifically, 
i) the accuracy are 1.9\% lower than the whole model if we remove attention, which means that the attention mechanism can fuse physical and semantic visual feature vectors better than simply concatenating, further proving our assumption that these features contribute to the task dynamically;
ii) When we remove the Bi-GRU from the pixel domain sub-network, the performance of MVNN reduces 1.8\%; when we further remove the branches, the performance drops by 4.3\%. Therefore, we can conclude that incorporating different levels of features and considering the dependencies between these features both help capture the semantic characteristics of visual contents.

\noindent
\textbf{Qualitative Analysis}

In the previous ablation experiments, we have intuitively evaluated the utility of multiple domains in improving the performance of MVNN. 
In this subsection, 
we qualitatively visualize the visual features learned by the frequency domain sub-network, the pixel domain sub-network and MVNN which considers both two domains on the testing set with t-SNE\cite{tsne}, as shown in Fig.~\ref{Fig:tsne}. The label for each image is real-news image or fake-news image.

From Fig.~\ref{Fig:tsne}, we can easily observe that the separability of the feature representations learned by these three networks can be sorted as: MVNN $>$ the pixel domain sub-network $>$ the frequency domain sub-network.
Specifically,
in the visualization graph of the frequency domain sub-network, the feature representations of positive and negative samples overlap a lot. 
This is because all images uploaded to social media platforms will be compressed, which covers the original compression or tampering traces to some extent, reducing the difference between fake-news images and real-news images in frequency domain.
For the pixel domain sub-network, it can learn discriminable features, but the learned features are still twisted together, especially in the middle part of Fig.~\ref{Fig:tsne-pixel}.
In comparison, in the visualization graph of MVNN, these is a relatively visible boundary between samples with different labels.
From above phenomena, we can conclude that: 
i) pixel domain is more effective than frequency domain in distinguishing fake-news images and real-news images; and
ii) frequency and pixel domains are complementary for detecting fake-news images, consequently MVNN which fuses the information of multiple domains can learn better and more distinctive feature representations, and thus achieves better performance than single domain sub-network.

\subsection{Application on Multimodal Fake News Detection}
In the previous parts, we have proven the effectiveness of MVNN in detecting fake news utilizing visual contents alone.
Considering that a complete news story consists of text and visual content, latest researches usually utilize multimodal information to detect fake news.
Following the convention, 
we compare the performance of different visual representations obtained from above visual modeling methods in fake news detection under the condition of fusing text information \textbf{(EQ3)}, as described in Tab.~\ref{tab:multimodality}.
We experiment with three state-of-the-art fusing methods as follows:
\begin{itemize}
	\item \textbf{attRNN}: Reference \cite{jin2017multimodal} proposes an innovative RNN with an attention mechanism for effectively fusing the textual, visual and social context features. 
	In detail, the neuron attention from the output of the LSTM is utilized when fusing with the visual features.
	In our experiments, we remove the part dealing with social context features. 
	\item \textbf{EANN}: Reference \cite{eannwang2018} designs an event adversarial neural network to learn event-invariant multimodal features through adversarial training techniques. It consists of a multimodal feature extractor, a fake news detector and an event discriminator. 
	\item \textbf{MVAE}: Reference \cite{mvae} utilizes a multimodal variational autoencoder trained jointly with a fake news detector to learn a shared representation of textual and visual information for fake news detection.
	MVAE is composed of an encoder, a decoder and a fake news detector. 
\end{itemize}

\begin{table}[b]
	\caption{The performance comparison of fake news detection in multi-modality.}
	\begin{center}
		\begin{tabular}{clcccc}
			\hline
			\multicolumn{2}{c}	{\textbf{Method}}&\textbf{Accuracy}&\textbf{Precision}&\textbf{Recall}&\textbf{F1}\\
			\hline
			\multirow{5}{*}{\textbf{attRNN}}&FF+LR & 0.735 & 0.801 & 0.665 & 0.727\\
			&Pre-trained VGG & 0.821 & 0.813 & 0.862 & 0.837 \\
			&Fine-tuned VGG & 0.849 & 0.888 &0.818 &0.852 \\
			&ConvAE & 0.816 & 0.848 & 0.796 & 0.821 \\
			&\textbf{MVNN} & \textbf{0.901} & \textbf{0.911} & \textbf{0.901} & \textbf{0.906}\\
			\hline
			\multirow{5}{*}{\textbf{EANN}}&FF+LR & 0.780 & 0.840 & 0.724 & 0.778 \\
			&Pre-trained VGG & 0.821 & 0.861 & 0.791& 0.824\\
			&Fine-tuned VGG & 0.841 & 0.883 & 0.807 & 0.843 \\
			&ConvAE & 0.823 & 0.863 & 0.794 & 0.827\\
			&\textbf{MVNN} & \textbf{0.897} & \textbf{0.930} & \textbf{0.872} & \textbf{0.900} \\
			\hline
			\multirow{5}{*}{\textbf{MVAE}}&FF+LR & 0.777 & 0.776 & 0.815 & 0.795 \\
			&Pre-trained VGG & 0.813 & 0.893 & 0.737 & 0.804 \\
			&Fine-tuned VGG & 0.832 & 0.875 & 0.798 & 0.835 \\
			&ConvAE & 0.827 & 0.831 & 0.847 & 0.839 \\
			&\textbf{MVNN} & \textbf{0.891} & \textbf{0.896} & \textbf{0.898} & \textbf{0.897} \\
			\hline
		\end{tabular}
		\label{tab:multimodality}
	\end{center}
\end{table}

Tab.~\ref{tab:multimodality} shows the similar trend as Tab.~\ref{tab:visual}.
From Tab.~\ref{tab:multimodality}, we observe that MVNN consistently outperforms other baselines by a large margin on all fusion methods. Specifically, MVNN exceeds compared methods by over 5.2\% in accuracy, which suggests that MVNN can easily replace existing methods to obtain the representations of visual contents, for achieving significant improvements in detecting fake news. 
Moreover, the performance of FF+LR in attRNN is obviously worse than EANN and MVAE, because attRNN can hardly utilize the semantic alignment between the text and the forensics features to fuse the textual and visual information.

\subsection{Case Studies}
In order to further illustrate the importance of multiple domains for fake news detection, we compare the results reported by MVNN and single domain sub-networks, and exhibit some successful examples which are captured by MVNN but missed by single domain sub-networks.

\begin{figure}
	[t]
	\centering
	\subfloat[]{
		\label{Fig:ffrequencycase1}
		\includegraphics[width=.48\linewidth,height=.4\linewidth]{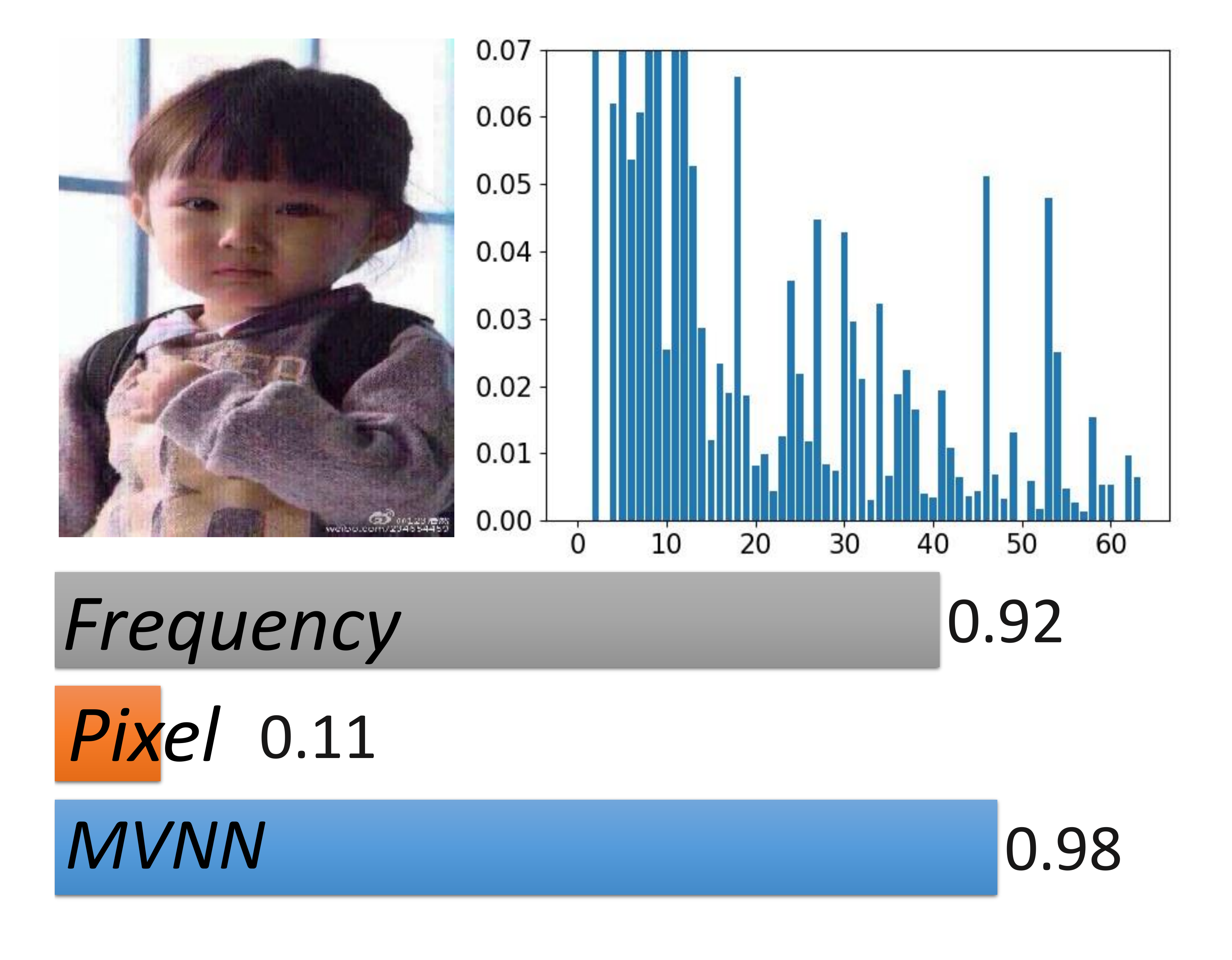}
	}
	\subfloat[]{
		\label{Fig:ffrequencycase2}
		\includegraphics[width=.52\linewidth,height=.4\linewidth]{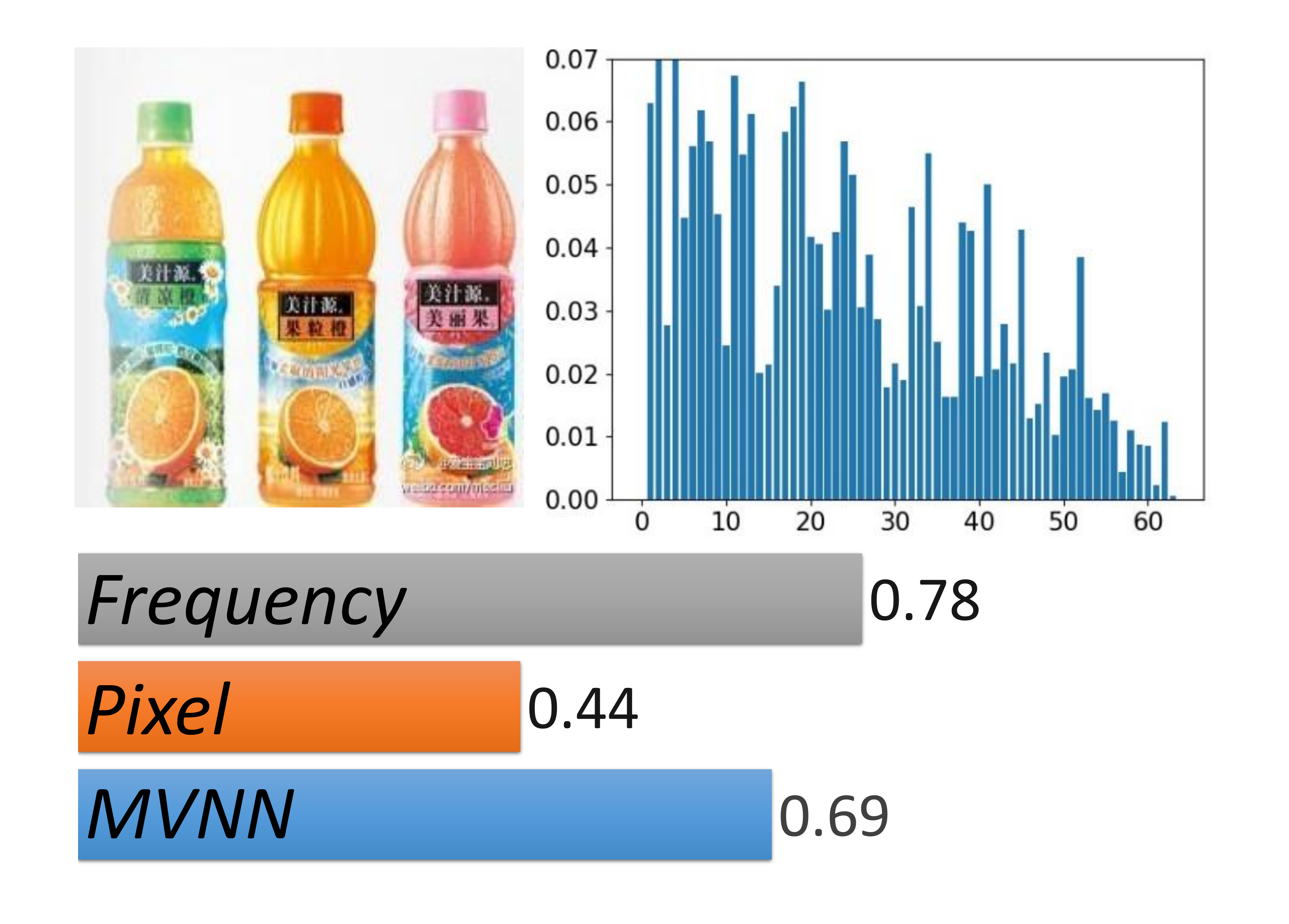}
	}\\
	\caption{Some fake-news images detected by the frequency domain sub-network and MVNN but missed by the pixel domain sub-network.}
	\label{Fig:ffrequencycase}
\end{figure}

Fig.~\ref{Fig:ffrequencycase} shows two fake-news images detected by the frequency domain sub-network and MVNN but missed by the pixel domain sub-network. 
Fig.~\ref{Fig:ffrequencycase1} was used to make up a fake news post about a lost child, and Fig.~\ref{Fig:ffrequencycase2} is an illustration of a fake news which says that juice drinks can cause leukemia in children.
The image in the upper left is the original image attached in the fake news post, and the histogram in the upper right shows the variance of DCT coefficients on 64 frequencies, 
which reflects the characteristics of this image in the frequency domain. In general, the larger the variance is, the heavier the image has been re-compressed.
From the perspective of semantics, the two images in Fig.~\ref{Fig:ffrequencycase} do not show evidence of fake news while their frequency histograms look quite suspicious, showing that they are very likely to be outdated images.
The scores obtained from MVNN and single domain sub-networks are listed in the bottom of each figure.
We can observe that these two examples are correctly classified by MVNN while the results would totally change without the frequency information, which further proves the importance of the frequency domain information in fake news detection.

\begin{figure}
	\centering
	\subfloat[]{
		\label{Fig:fpixelcase1}
		\includegraphics[width=.48\linewidth,height=.4\linewidth]{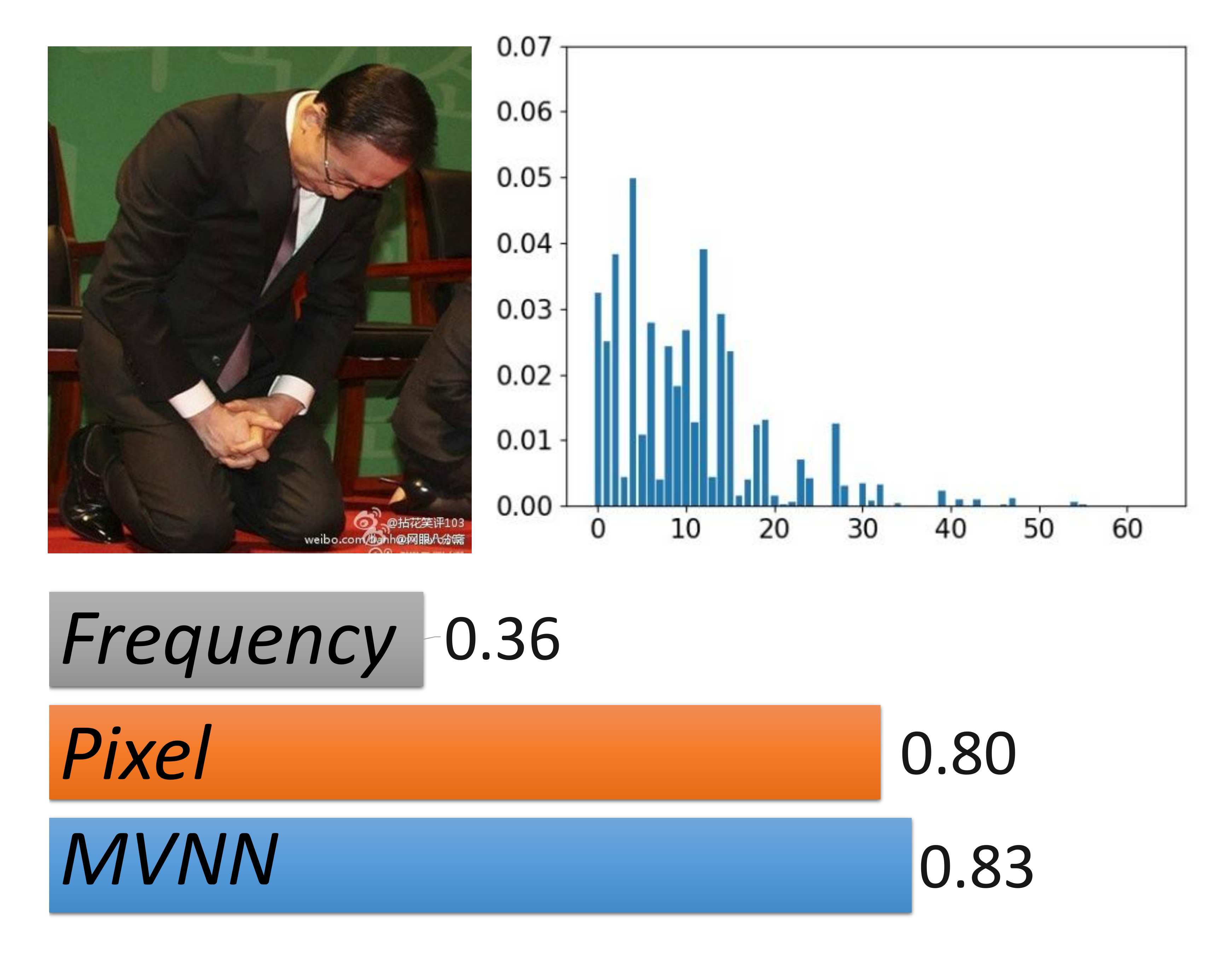}
	}
	\subfloat[]{
		\label{Fig:fpixelcase2}
		\includegraphics[width=.52\linewidth,height=.4\linewidth]{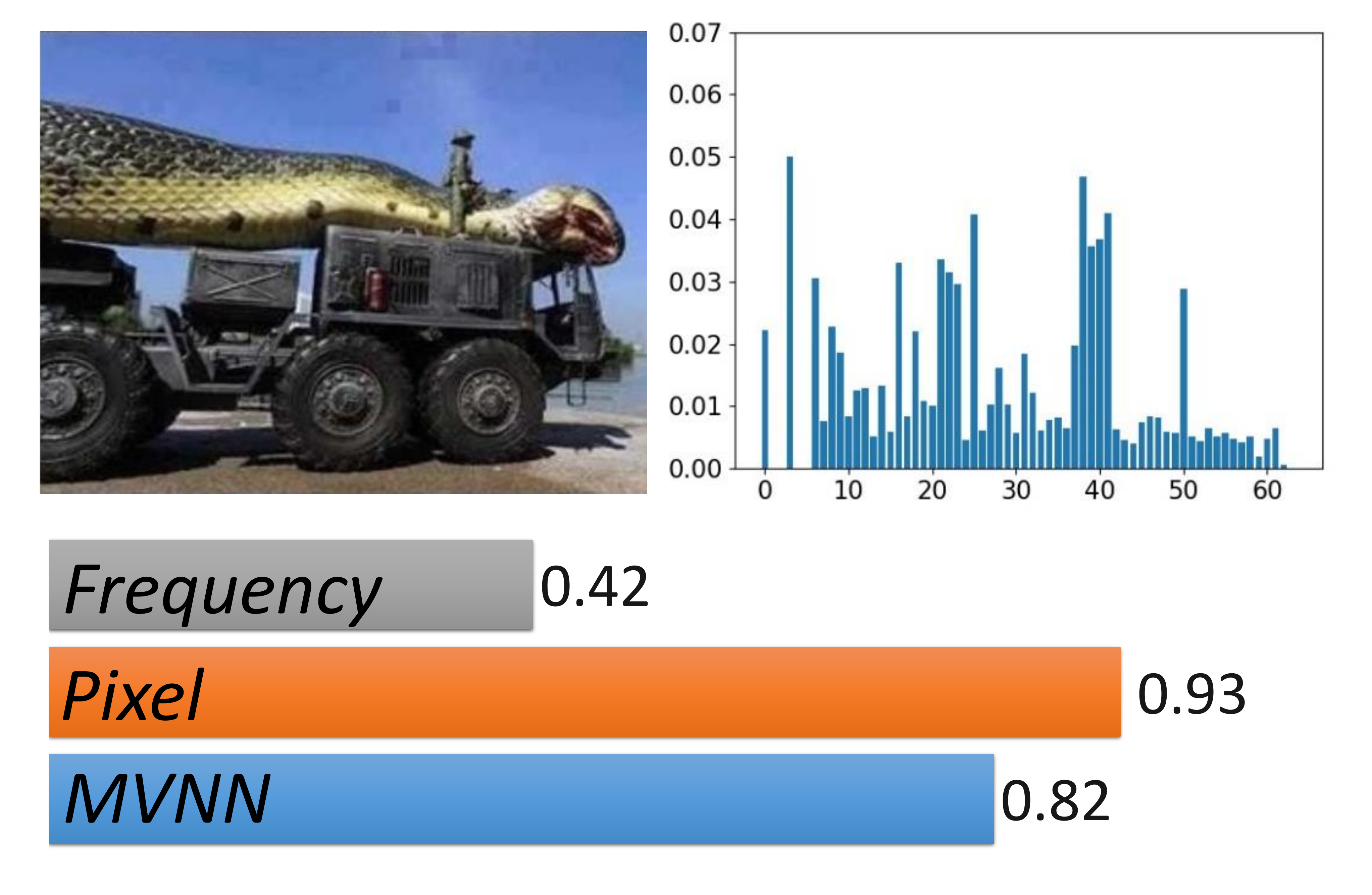}
	}\\
	\caption{Some fake-news images detected by the pixel domain sub-network and MVNN but missed by the frequency domain sub-network.}
	\label{Fig:fpixelcase}
\end{figure}

Fig.~\ref{Fig:fpixelcase} shows another two fake-news images detected by the pixel domain sub-network and MVNN but missed by the frequency domain sub-network. 
Fig.~\ref{Fig:fpixelcase1} tells a fake news that the South Korean President Lee Myung-bak knelt down to apologize to the nation for the sexual assault of a seven-year-old girl.
In fact, he was just praying.
Fig.~\ref{Fig:fpixelcase2} is used to fabricate a fake news story about a huge snake which is just a mock-up. 
We can see that the two frequency histograms in Fig.~\ref{Fig:fpixelcase} show little evidence of fake news, however, the image contents are eye-attracting and rather dubious. 
By combining the information of the frequency and pixel domain, MVNN can easily detect that this is fake-news image with high confidence.
Therefore, the information of the pixel domain is also important for detecting fake news.


\section{Conclusions and Future Work}
In this paper, we propose a novel framework MVNN to model the visual contents for fake news detection, which takes advantage of the visual information of frequency and pixel domains to effectively capture and fuse the characteristics of fake-news images at physical and semantic levels.    
Experiments conducted on Weibo dataset validate the effectiveness of MVNN, and further prove the importance of multiple domains in fake news detection.

There are several future works that need further investigations. 
First, 
MVNN is a general framework for extracting effective visual representations to detect fake news, which is not limited to platforms. 
In this paper, we only evaluate the proposed model on Weibo data due to the limited distinctive images in the existing Twitter multimedia dataset. 
In future work, we can construct a larger multimedia dataset from Twitter platform and explore the generalization capacity of the proposed model on different datasets.
Further, we can compare the similarities and differences between the visual contents of Weibo and Twitter data.
Second, although there are already many studies focusing on fusing multimodal information for fake news detection, it is still a challenging problem which needs further investigation.
For example, we can use the semantic alignment between images and text to explore the role of different modalities.
At last, how to explain the decisions made by existing models based on multimodal information is worth considering, since it can help us further understand and defend fake news.

\section*{Acknowledgment}

\bibliographystyle{IEEEtran}
\bibliography{sample-base}

\end{document}